\documentclass[onecolumn]{revtex4}

\usepackage{graphicx}
\usepackage{dcolumn}
\usepackage{bm}

\input epsf

\begin{document}

\title{\Large Emergent Universe in Brane World Scenario with
Schwarzschild-de Sitter Bulk}

\author{\bf Asit~Banerjee$^1$, Tanwi~Bandyopadhyay$^2$ and
~Subenoy~Chakraborty$^2$\footnote{schakraborty@math.jdvu.ac.in}}

\affiliation{$^1$Department of Physics, Jadavpur University,
Kolkata-32, India.\\ $^2$Department of Mathematics, Jadavpur
University, Kolkata-32, India.}

\date{\today}

\begin{abstract}
A model of an emergent universe is obtained in brane world. Here the
bulk energy is in the form of cosmological constant, while the brane
consists of a fluid satisfying an equation of state of the form
$p_{b}=\frac{1}{3}~\rho_{b}$, which is effectively a radiation
equation of state at high energies. It is shown that with the
positive bulk cosmological constant, one of our models represents an
emergent universe.
\end{abstract}

\maketitle

The search for singularity free inflationary models in the
context of Classical General Relativity has recently led to the
development of emergent universes.\\

In 1967 Harrison [1] obtained a model of the closed universe
containing radiation, which approaches the state of an Einstein
static model asymptotically, i.e, as $t\rightarrow-\infty$. This
kind of model has so far been discovered subsequently by several
workers in the recent past such as that of Ellis and Maartens
[2], Ellis et al. [3]. They obtained closed universes with a
minimally coupled scalar field $\phi$ with a special form for
self interacting potential and possibly some ordinary matter with
equation of state $p=\omega\rho$ where
$-\frac{1}{3}\leq\omega\leq1$. However, exact analytic solutions
were not presented in these models, although their behaviour
alike that of an emergent universe was highlighted. An emergent
universe is a model universe in which there is no timelike
singularity, is ever existing and having almost static behaviour
in the infinite past ($t\rightarrow-\infty$) as is mentioned
earlier. The model eventually evolves into an inflationary stage.
In fact, the emergent universe scenario can be said to be a
modern version and extension of the original Lemaitre-Eddington
universe. Mukherjee et al. [4] obtained solutions for Starobinsky
model for flat FRW space time and studied the features of an
emergent universe. Very recently, a general framework for an
emergent universe model has been proposed by Mukherjee et al. [5]
using an adhoc equation of state connecting the pressure and
density. However, these solutions require exotic matter in many
cases. Such models are appealing since they provide specific
examples of non singular (geodesically complete) inflationary
universes. Further, it has been proposed that entropy
considerations favor the Einstein static state as the initial
state for our universe [6].\\

The emergent universe models mentioned above for four dimensional
space-time assume features like positive spatial curvature,
minimally coupled scalar field or exotic matter. So far, we have not
noticed any emergent universe model in the brane with the background
of a bulk with cosmological constant. In the present work, we
consider a perfect fluid satisfying an equation of state

\begin{equation}
p_{b}=\frac{1}{3}~\rho_{b}
\end{equation}

The evolution of the brane world models in which we live, in the
special case, interestingly shows the feature of an emergent
universe behaviour. The models described here are not only
homogeneous and isotropic at large scale but also are spatially
flat.\\

The brane energy density consists of two parts, one due to the
ordinary energy density and the other due to the so called brane
tension. The brane energy density and pressure are therefore given
respectively as $\rho_{b}=\rho+\sigma$ and $p_{b}=p-\sigma$.
Initially at large density $\rho>>\sigma$ so that
$p\approx\frac{1}{3}\rho$, which is effectively a radiation equation
of state. On the other hand, as we show in what follows that at late
stage, that is at $a_{0}\rightarrow\infty$ and $\rho\rightarrow0$,
the brane inflates. There are not so many exact cosmological
solutions in the brane because the energy density of the brane
appears quadratically in the modified Friedmann equations instead of
its linear behaviour as in the usual equations. We believe that
particularly the emergent universe model in the brane of our present
paper is the first of its kind in the existing
literature.\\

The geometry of the five dimensional bulk is assumed to be
characterized by the space time metric of the form

\begin{equation}
ds^{2}=-n^{2}(t,y)dt^{2}+a^{2}(t,y)\delta_{ij}dx^{i}dx^{j}+b^{2}(t,y)dy^{2}
\end{equation}

where $y$ is the fifth coordinate and the hypersurface $y=0$ is
identified as the world volume of the brane that forms our universe.
For simplicity, we choose the usual spatial section of the brane to
be flat. Now following Bin\'{e}truy et al. [7,~8], the energy
conservation equation on the brane reads

\begin{equation}
\dot{\rho_{b}}+3(\rho_{b}+p_{b})\frac{\dot{a_{0}}}{a_{0}}=0
\end{equation}

which integrating once (using the equation of state (1)) gives

\begin{equation}
\rho_{b}=\frac{\rho_{0}}{a_{0}^{4}}
\end{equation}

($\rho_{0}$, an arbitrary integration constant)\\

Using this form for $\rho_{b}$, the generalized Friedmann type
equations take the form (see eqns. (45) and (46) in ref. [9])

\begin{equation}
\frac{\dot{a_{0}^{2}}}{a_{0}^{2}}=\frac{\kappa^{4}\rho_{b}^{2}}{36}
+\frac{\kappa^{2}\Lambda_{5}}{6}+\frac{C}{a_{0}^{4}}
\end{equation}

and

\begin{equation}
\ddot{a_{0}}=-\frac{\kappa^{4}}{12}\frac{\rho_{0}^{2}}{a_{0}^{7}}
+\frac{\kappa^{2}\Lambda_{5}}{6}~a_{0}-\frac{C}{a_{0}^{3}}
\end{equation}

Here $C$ appears as an integration constant.\\

We now proceed to solve the equation (5). The integration leads to

\begin{equation}
\frac{1}{4}\int\frac{du}{\sqrt{bu^{2}+Cu+d}}=\pm~(t-t_{0})
\end{equation}

with $u=a_{0}^{4},~b=\frac{\kappa^{2}\Lambda_{5}}{6},~d=\frac{\kappa^{4}\rho_{0}^{2}}{36}$.\\

The explicit solutions for $b>0$ are then given by

\begin{equation}
a_{0}^{4}=\left\{
\begin{array}{ll}
\frac{\sqrt{4bd-C^{2}}}{2b}~Sinh\left[4\sqrt{b}~(t-t_{0})\right]-\frac{C}{2b},
~~~~~~~~~~~~~~~~(\text{when $4bd>C^{2}$})\\\\
\frac{\sqrt{C^{2}-4bd}}{2b}~Cosh\left[4\sqrt{b}~(t-t_{0})\right]-\frac{C}{2b},
~~~~~~~~~~~~~~~~~~(\text{when $4bd<C^{2}$})\\\\
\frac{1}{2b}\left[e^{4\sqrt{b}~(t-t_{0})}\right]-\frac{C}{2b},
~~~~~~~~~~~~~~~~~~~~~~~~~~~~~~~~~~~(\text{when $4bd=C^{2}$})
\end{array}
\right.
\end{equation}

We note that $b>0$ implies positive bulk cosmological constant
($\Lambda_{5}>0$). Now for $C>0$, all the above solutions start from
big bang singularity and expand indefinitely as
$t\rightarrow\infty$. However, for $C<0$, the behaviour of the third
solution is quite different. It is then a singularity free solution
which starts with finite $a_{0}$ at $t\rightarrow-\infty$, where
both $\dot{a_{0}}$ and $\ddot{a_{0}}$ vanish and subsequently shows
exponential expansion. The brane model corresponding to this
solution is termed as an emergent universe [1-3] in the brane world
scenario.\\

We note that in the equation (5), the term including $C$ is called
the black radiation term, contributed from the bulk Weyl tensor. This
parameter $C$ arises from the analysis of the bulk [8,~10-12].\\

The five dimensional bulk space-time with flat spatial section may
be written in the form (see Ida [13])

\begin{equation}
ds^{2}=-h(a)dt^{2}+\frac{1}{h(a)}~da^{2}+a^{2}\left[d\chi^{2}
+\chi^{2}d\Omega^{2}\right]
\end{equation}

We may locate the three brane in the form of $t=t(\tau)$,
$a=a(\tau)$ parametrized by the proper time $\tau$ on the brane,
there the induced metric of three brane will be given by

\begin{equation}
ds_{(4)}^{2}=-d\tau^{2}+a^{2}(\tau)\left[d\chi^{2}
+\chi^{2}d\Omega^{2}\right]
\end{equation}

where $\tau$ and $a(\tau)$ correspond to the cosmic time and scale
factor respectively for spatially flat Friedmann-Robertson-Walker
universe.\\

The solution of the bulk field equations
$R_{\mu\nu}^{(5)}=\Lambda_{5}~g_{\mu\nu}^{(5)}$ is shown clearly in
[13] that for $\Lambda_{5}>0$ and $C<0$, the bulk space-time becomes
Schwarzschild-de Sitter type and the black hole horizon exists,
where $h(a)=0$. In fact, $C$ plays the role of the black hole mass.
It is interesting to note that because of the positive magnitude of
the bulk cosmological constant ($\Lambda_{5}>0$), the naked
singularity can be avoided even if $C$ is chosen to be negative in
the emergent universe model and the bulk background has a horizon
for the bulk singularity. We further note from equation (6) that,
with $\Lambda_{5}>0$ and $C<0$, the emergent universe model evolves
into an accelerated expansion at the late stage.\\

Moreover, for negative cosmological constant (i.e, $b<0$), the
solution can be written as

\begin{equation}
a_{0}^{4}=\frac{C}{2|b|}+\frac{\sqrt{4|b|d+C^{2}}}{2|b|}
\begin{array}{c}
  Sin \\
  or \\
  Cos
\end{array}
[4\sqrt{|b|}~(t-t_{0})]
\end{equation}

which has familiar behaviour and is not of much interest in the
present context.\\

We shall now discuss the properties of that solution in equations
(8) representing a model for emergent universe. Asymptotically in
the past (i.e, at $t\rightarrow-\infty$) the scale factor $a_{0}$
has the constant value $(\frac{|C|}{2b})^{1/4}$ and using equation
(4), the matter density has initially the constant value

\begin{equation}
\rho_{bi}=\left(\frac{6\Lambda_{5}}{\kappa^{2}}\right)^{1/2}
\end{equation}

Further, it should be noted that equation (5) may also be written as

\begin{equation}
\frac{\dot{a_{0}^{2}}}{a_{0}^{2}}=\frac{\kappa^{4}\rho^{2}}{36}
+\frac{\kappa^{4}\sigma\rho}{18}+\frac{\kappa^{2}\Lambda_{4}}{6}
\end{equation}

where $\Lambda_{4}=\Lambda_{5}+\frac{\kappa^{2}\sigma^{2}}{6}$, is
the effective four dimensional cosmological constant and is non-zero
positive constant due to positive $\Lambda_{5}$.\\

In the present work, we have followed the procedure adopted by
Bin\'{e}truy et al. [7-8], in which the Randall-Sundrum spacetime was
generalized to allow for time dependent cosmological expansion. We
are mainly concerned with the derivation of the four dimensional
brane cosmological equations using the $Z_{2}$ symmetry at the
boundary. Using Gauss-Codazzi formulation with Israel's junction
conditions [14] on the brane, Roy Maartens [15] and Shiromizu et al.
[16] have formulated a pure brane Einstein equations, where the bulk
geometry reflects through the projected Weyl tensor term. It should
be noted that this approach yields exactly the same results as [7,~8]
for FRW branes.\\

{\bf Acknowledgement}:\\

The authors are thankful to the editor for his useful suggestions in
revising the manuscript. Also, TB is thankful to CSIR, Govt. of
India, for
awarding JRF.\\


\begin{thebibliography}{2}

\bibitem{Harrison} E. R. Harrison, {\it Mon.Not.R.Astron.Soc.}
{\bf 137}, 69 (1967).\\

\bibitem{EM} G. F. R. Ellis and R. Maartens, {\it
Class.Quant.Grav.} {\bf 21}, 223 (2004), {\it gr-qc}/0211082
(2003).\\

\bibitem{EMT} G. F. R. Ellis, J. Murugan and C. G. Tsagas, {\it
Class.Quant.Grav.} {\bf 21}, 233 (2004), {\it gr-qc}/0307112
(2003).\\

\bibitem{Mukherjee} S. Mukherjee, B. C. Paul, S. D. Maharaj and A.
Beesham, {\it gr-qc}/0505103 (2005).\\

\bibitem{MPDMB} S. Mukherjee, B. C. Paul, N. K. Dadhich, S. D. Maharaj and A.
Beesham, {\it Class.Quant.Grav.} {\bf 23}, 6927 (2006).\\

\bibitem{Gibbons} G. W. Gibbons, {\it Nucl.Phys.B} {\bf 292}, 784
(1988); ibid {\bf 310}, 636 (1988).\\

\bibitem{BDL} P. Bin\'{e}truy, C. Deffayet and David Langlois,
{\it Nucl.Phys.B} {\bf 565}, 269 (2000).\\

\bibitem{BDEL} P. Bin\'{e}truy, C. Deffayet U. Ellwanger and D.
Langlois, {\it Phys.Lett.B} {\bf 477}, 285 (2000), {\it hep-th}/9910219 (2000).\\

\bibitem{BCB} T. Bandyopadhyay, S. Chakraborty and A. Banerjee,
{\it gr-qc}/0609067 (2006).\\

\bibitem{Mukohyama} S. Mukohyama, T. Shiromizu and K. Maeda, {\it
Phys.Rev.D} {\bf 62}, 024028 (2000).\\

\bibitem{Creek} S. Creek, R. Gregory, P. Kanti and B. Mistry, {\it
Class.Quant.Grav.} {\bf 23}, 6633 (2006).\\

\bibitem{VDB} P. Brax and C. van de Bruck, {\it Class.Quant.Grav.}
{\bf 20}, R201 (2003).\\

\bibitem{Ida} D. Ida, {\it JHEP} {\bf 0009}, 014 (2000).\\

\bibitem{Israel} W. Israel, {\it Nuovo Cim.} {\bf 44B}, 1 (1966).\\

\bibitem{Roy} Roy Maartens, {\it Living.Rev.Rel.} {\bf 7}, 7 (2004).\\

\bibitem{SMS} T. Shiromizu, K. Maeda and M. Sasaki, {\it Phys.Rev.D}
{\bf 62}, 024012 (2000).\\


\end{thebibliography}
\end{document}